\documentclass[12pt]{iopart}
\usepackage{iopams}
\usepackage{graphicx}
\usepackage{verbatim}
\begin{document}

\title[P. Chandra et al.]{TeV  Gamma-ray  Observations of   Markarian 421  using TACTIC   during  2009-10}

\author{  P. Chandra,  R.C. Rannot, K.K. Yadav, A.K. Tickoo, K. K. Singh, K. Chanchalani, M. Kothari,  
 N.K. Agarwal, A. Goyal, H.C. Goyal,  S. Kotwal, 
N. Kumar, P. Marandi, K.Venugopal, C.K. Bhat, N. Bhatt, S. Bhattacharya, C. Borwankar, N. Chouhan, V.K. Dhar, S.R. Kaul, S.K. Koul, M.K. Koul, R. Koul, A.K. Mitra, S. Sahayanathan, M. Sharma}

\address {   Astrophysical Sciences Division, Bhabha Atomic Research Centre, \\
Mumbai - 400 085, India.}

\ead{rcrannot@barc.gov.in}
\begin{abstract}
 We have observed  the  blazar   Markarian 421  with the TACTIC $\gamma$-ray telescope at Mt. Abu,  India, from  22 November 2009 to 16 May 2010   for 265 hours. Detailed   analysis of  the  data so recorded  revealed  presence of a  TeV $\gamma$-ray signal   with a statistical significance of 12.12$\sigma$  at $E_{\gamma}\geq$ 1 TeV. We have    estimated  the  time averaged  differential energy spectrum of  the source   in the energy range 1.0 - 16.44 TeV.  The spectrum  fits well with the power law function of the form ($dF/dE=f_0 E^{-\Gamma}$) with  $f_0=(1.39\pm0.239)\times 10^{-11}cm^{-2}s^{-1}TeV^{-1}$ and  $\Gamma=2.31\pm0.14$.
\end{abstract}
\section{Introduction} 
Markarian 421 falls in the blazar category of Active Galactic Nuclei (AGN) which are the most extreme and powerful variable sources of  photons with energies ranging from the radio to the $\gamma$-ray regimes.  Blazars   are believed to have their jets more aligned towards us  as compared to  any other class of radio loud AGN, are quite bright, have irregular  amplitude variability in all accessible spectral bands and  a core-dominated radio morphology with flat  radio spectra, which join smoothly to the infra-red (IR), optical and ultra-violet spectra.  % The corresponding fluxes in different electromagnetic bands   exhibit high and variable polarization. With more observations with a variety of ground and satellite based instruments  a variety of features of such objects are becoming prominent  and many different models can explain their observed properties \cite{boettcher07,sambruna07}.
The observed   high variability and  broadband emission from such objects make long-term multiwavelength observations of blazars  very important for  understanding their emission mechanisms and other jet related  properties\cite{horan09, urry95}.
The Spectral Energy Distributions (SEDs) of these violent objects have  double-peaked structure in the $\nu$F$_{\nu}$ versus frequency plot.%Both peaks are found to vary, often both in intensity and  peak frequency as the activity level of the blazar changes.
The first peak is usually referred to as the synchrotron peak,  both in leptonic and hadronic models for blazar emission. Further, it is  generally believed  to be  the result of incoherent synchrotron emission from relativistic electrons and positrons, which are conjectured to be present in the magnetic fields of the jet. The origin of the second peak, usually referred to as the inverse-Compton peak, is less well determined. In synchrotron self-Compton (SSC) models\cite{Bloom96,Konigl81,Maraschi92,Sikora01} it is assumed that the synchrotron photons are up-scattered to higher energies by the electrons while in external Compton (EC) models\cite{Blandford95,Dermer97,Ghisellini96,Wagner95}, these seed photons can come from the accretion disk,  broad-line region,  torus, local infrared background,  cosmic microwave background,  ambient photons from the central accretion flow or some combination of these sources.
In the framework of hadronic models
% have also been invoked to explain the broadband spectra of blazars
\cite{ah00,man93,muc03}, %it is  proposed that
the X-ray to $\gamma$-ray emission is believed to be due to synchrotron radiation from protons accelerated in highly magnetized compact regions of the jet. In another scenario it is proposed that the proton-proton collisions, either within the jet itself or between the jet and ambient clouds, give rise to neutral pions which  decay to gamma rays \cite{Beall99,Dar97,Pohl00}.

The blazar Mrk 421(z=0.031)  was the first extragalactic source detected at TeV energies in 1992 using imaging atmospheric Cherenkov telescopes \cite{pet96,pun92}. 
The source has been  observed  extensively by various groups since then 
\cite{Abdo11, ah03, ah05, al06,Bart11,boo02,chandra10,Mart11,nico11,Ong10, pir01, sm06,ya07,zw97}.
These observations have shown that the TeV $\gamma$-ray emission from Mrk 421 is highly variable with variations of more than one order of magnitude and occasional flaring doubling time  as short as 15 mins \cite{ah02,gai96}. 
Since its detection in the TeV energy range, Mrk 421 has also been  the target of several multiwavelength observation campaigns \cite{Abdo11,ale10,Bart11,blaz05,buc96,don09,tak96,tak00,wag09}. 
Several groups have also determined the energy spectrum  of Mrk 421 at various  flux levels %\cite{ah05,al06,zw97,acc09,ah99} \cite{ah02,pir01,ya07,he11,Mart11} 
and some of these results are cited and discussed in section 6. 
%The recent results of these studies \cite{ah03,ah05,kren02,pir01} suggest that  the spectrum is compatible with a power law combined with an exponential  cutoff. 
%It has also been reported  that the spectrum hardens as the flux increases \cite{ah03,kren02}, either because of an increase in the cutoff energy or a change in the spectral index itself.
Differences in the energy spectrum of Mrk 421 and Mrk 501  have also been addressed to understand the  $\gamma$-ray production mechanisms of these objects and  absorption effects at the source or in the intergalactic medium due to interaction of $\gamma$-rays with the Extragalactic Background Light(EBL) \cite{dw05}. In this paper, we present TACTIC results obtained from  our observation campaign  on this source during the period 2009-10. 
%It may be noted that the  TACTIC 2008 observations  probes the source spectrum  just before the June 2008 flare recorded by the MAGIC, VERITAS, ARGO-YBJ and \textit{AGILE }detectors \cite{don09,wag09,aie10,don09}. 
%In addition we  also compare the TACTIC  TeV light curves with those obtained with the Rossi X-Ray Timing Explorer's All-Sky Monitor (RXTE/ASM) \cite{ASM}, \textit{Swift} Burst Alert Telescope (BAT) \cite{swift} and Fermi (LAT) satellite based experiments which are sensitive at  X-ray energies. 

\section{ TACTIC telescope}
TACTIC (TeV Atmospheric Cherenkov Telescope with Imaging Camera) $\gamma$-ray telescope is  located  at Mt. Abu ( 24.6$^\circ$ N, 72.7$^\circ$ E, 1300m asl), Rajasthan, India.  It deploys a F/1 type tracking light collector of $\sim$9.5 m$^2$ area which is  made up of 34   0.6-m-diameter front-coated spherical glass facets which have been prealigned to produce an on-axis spot of $\sim$ 0.3$^\circ$ diameter at the focal plane. The telescope has a 349 photomultiplier tubes (ETL 9083UVB) -based imaging camera  with a uniform pixel resolution of $\sim$0.3$^\circ$ and a field-of-view of $\sim$6$^\circ$x6$^\circ$ to record images of atmospheric Cherenkov events. Data used in the present work  have been collected with the inner 225 pixels and the innermost 121 pixels were used for generating the event trigger. The trigger is based on the 3NCT (Nearest Neighbour Non-Collinear Triplets) logic \cite{Kaul03}. The safe anode current ($\leq$ 3 $\mu$A) operation of the photomultipliers (PMT) has been ensured by implementing a gain control algorithm \cite{Bhatt01}. The data acquisition and control system of the telescope \cite{ya04} have been designed around a network of PCs with the QNX (version 4.25) real-time operating system. The triggered events are digitized by CAMAC based 12-bit Charge to Digital Converters (CDC) which have a full scale range of 600 pC. The single pixel threshold was set to $\geq$ 14 photoelectrons(pe). The detailed description  of the telescope related hardware and software  has been given in   \cite{Koul07}.  The telescope is sensitive to $\gamma$-rays above 1 TeV and can detect the Crab Nebula at 5$\sigma$ significance level in 25 hours of  observation.  
%We have already reported TACTIC observational results on Mrk 421, Mrk 501, 1ES2344+514 and H1426+428 \cite{go07,go08,ya07,ya09}.  
\section{Mrk 421 Observations}
\begin{table}[!htp]
\caption{ Details of   Mrk421 observations using TACTIC  during the period 2009-10.}
\label{tab:obser}
\begin{center}

\begin{tabular}{|c|c|c|c|c|}
\hline
Year  &  Month & Observation dates & Total data (hrs.) & Data Selected (hrs.)\\ 
\hline
2009 &  Nov. & 22-27 &   5.6      &     5.6       \\  \hline
2009&  Dec. & 13,15-16,18-19,21,23-27 &   21.1      &       16.9     \\  \hline
2010&  Jan. & 10,13-24 &   47.8      &    41.0        \\  \hline
2010 &  Feb. & 6,10-20,22-23 &   54.2      &    50.2        \\  \hline
2010 &  Mar. & 7-22 &    60.5     &     59.4       \\  \hline
2010 &  Apr. & 2,4-7,9-18 &   50.2      &   38.5         \\  \hline
2010 &  May & 2,4,6-13,15-16 &   26.0      &  18.3         \\  \hline
2009-2010 &  Nov.-May &   &   265.4      &     229.9      \\  \hline
\end{tabular}
\newline\newline
\end{center}

\end{table} 
\vspace{1.0cm}
Very High Energy(VHE)  $\gamma$-ray observations of Mrk 421 presented here were made  with       the  TACTIC $\gamma$-ray telescope during   22 November 2009 to 16 May 2010   for 265.4 hours.  The observations were carried out in tracking mode, where the source is tracked continuously without taking  off- source data \cite{Quinn96}. This  mode  improves the chances of recording possible flaring activity from the candidate source direction. Details of these  observations with the TACTIC $\gamma$ -ray telescope during 2009-10 are given in Table \ref{tab:obser}. We have used the standard data quality checks  to evaluate the overall system behaviour and the general quality of the recorded data, including conformity of the Prompt Coincidence (PC) rates  with the expected zenith angle trend, compatibility of the arrival times of PC events with the Poissonian statistics and the steady behaviour of the chance coincidence rate with time \cite{Weekes89}. After applying these cuts, we have  selected good quality data sets  as per
details  given in  Table \ref{tab:obser}.
%which amounted to 235.9 hours of clean observations. 

\section{Data Analysis and Results} 
 The  analysis of  data  recorded  by the ground based atmospheric Cherenkov    gamma-ray telescopes,   involves  a number of steps including  filtering of  night sky background light, accounting for the differences in the relative gains of the PMTs and finding Cherenkov image boundaries, image parameterization,  event selection, energy reconstruction  etc. We have characterized each Cherenkov image  using a moment analysis methodology given in\cite{ Hillas85, Reynolds93, Weekes89,acc09, ah06, al08}, wherein the roughly elliptical shape of the image is described by the LENGTH and  WIDTH parameters and its location and orientation within the telescope field of view are given by the DISTANCE  and ALPHA($\alpha$) parameters respectively. In addition, the two highest amplitude   signals recorded by the PMTs (max1, max2) and the amount of light in the image (SIZE) are also obtained for each recorded image. Further,  another parameter   FRAC2  which is  the addition of two largest amplitude signals recorded by PMTs divided by the image size, is also estimated\cite{Weekes89}.
 The standard Dynamic Supercuts \cite{Hillas98,Mohanty98} procedure is then used to separate $\gamma$-ray like images from the overwhelming background of cosmic-rays.  This procedure uses the image shape parameters LENGTH and WIDTH   as a function of the image SIZE so that energy dependence of these parameters can be taken into account. The $\gamma$-ray selection criteria used in this analysis is  given in  Table \ref{tab:cuts} and   has been obtained on the basis of  Monte Carlo simulations carried out for the TACTIC telescope \cite{dhar09,Koul11}. 
\vspace{0.3cm}
\begin{table}[h]
\caption{ Dynamic Supercuts  selection  criteria used for analyzing the TACTIC data.}
\label{tab:cuts}
\vspace{0.3cm}
\begin{center}
\begin{tabular}{|c|c|}
\hline 
\hline
Parameters  &  Cuts Value \\ \hline
LENGTH (L)  &  0.11$\le$  L $\le$ (0.235 + 0.0265 $\times$lnS)$^{0}$ \\ \hline
WIDTH (W)  &  0.065$\le$  W $\le$ (0.085 + 0.0120 $\times$lnS)$^{0}$ \\ \hline
DISTANCE (D)  &  0.5$\le$  D $\le$ (1.27 Cos$^{0.88}$ln$\Theta$)$^{0}$ ($\Theta$=zenith angle) \\ \hline
SIZE (S)  &  S $\ge$  485 dc (6.5 digital counts = 1.0pe)\\ \hline
ALPHA ($\alpha$)  &  $\alpha$ $\le$ 18$^{0}$ \\ \hline
FRAC2 (F2)  &  F2 $\ge$  0.38\\ \hline
\hline
\end{tabular}
\end{center}
\end{table}  
\vspace{0.4cm}
\\ 
 VHE $\gamma$-ray signal  processing   is done  by using   the histogram of the $\alpha$  parameter (defined as the angle between the major axis of the image and the line between the image centroid and camera center, when the source is aligned along the optical axis of the telescope)  after applying the set of  image cuts given in Table \ref{tab:cuts}.  The distribution of  the $\alpha$  parameter  is expected to be flat for the isotropic background of cosmic ray events, whereas for the $\gamma$-ray signal  events, the distribution is expected to show a peak at smaller $\alpha$ values. This range for the TACTIC  telescope is  $\alpha$ $\leq$ 18$^{\circ}$. The contribution of the background events is estimated from a reasonably flat $\alpha$ region  of  27$^{\circ}$ $\leq$ $\alpha$ $\leq$ 81$^{\circ}$. The number of $\gamma$-ray events is then calculated by subtracting the expected number of background events, calculated on the basis of the background region \cite{Catanese98}, from the $\gamma$-ray domain events. The reason for not including the $\alpha$ bins 18$^{\circ}$-27$^{\circ}$ and 81$^{\circ}$-90$^{\circ}$ in the background region is to ensure that the background level is not overestimated because of a possible spill over of $\gamma$-ray events  in the third bin  and the truncation of the Cherenkov images recorded at the boundary of the imaging camera \cite{Catanese98} in the last bin. The significance of the excess events is calculated by using the maximum likelihood ratio method of Li and Ma \cite{LiMa83}.

\begin{figure}
\begin{center}
\mbox{\hspace{0cm}\includegraphics[scale=0.62,angle=90]{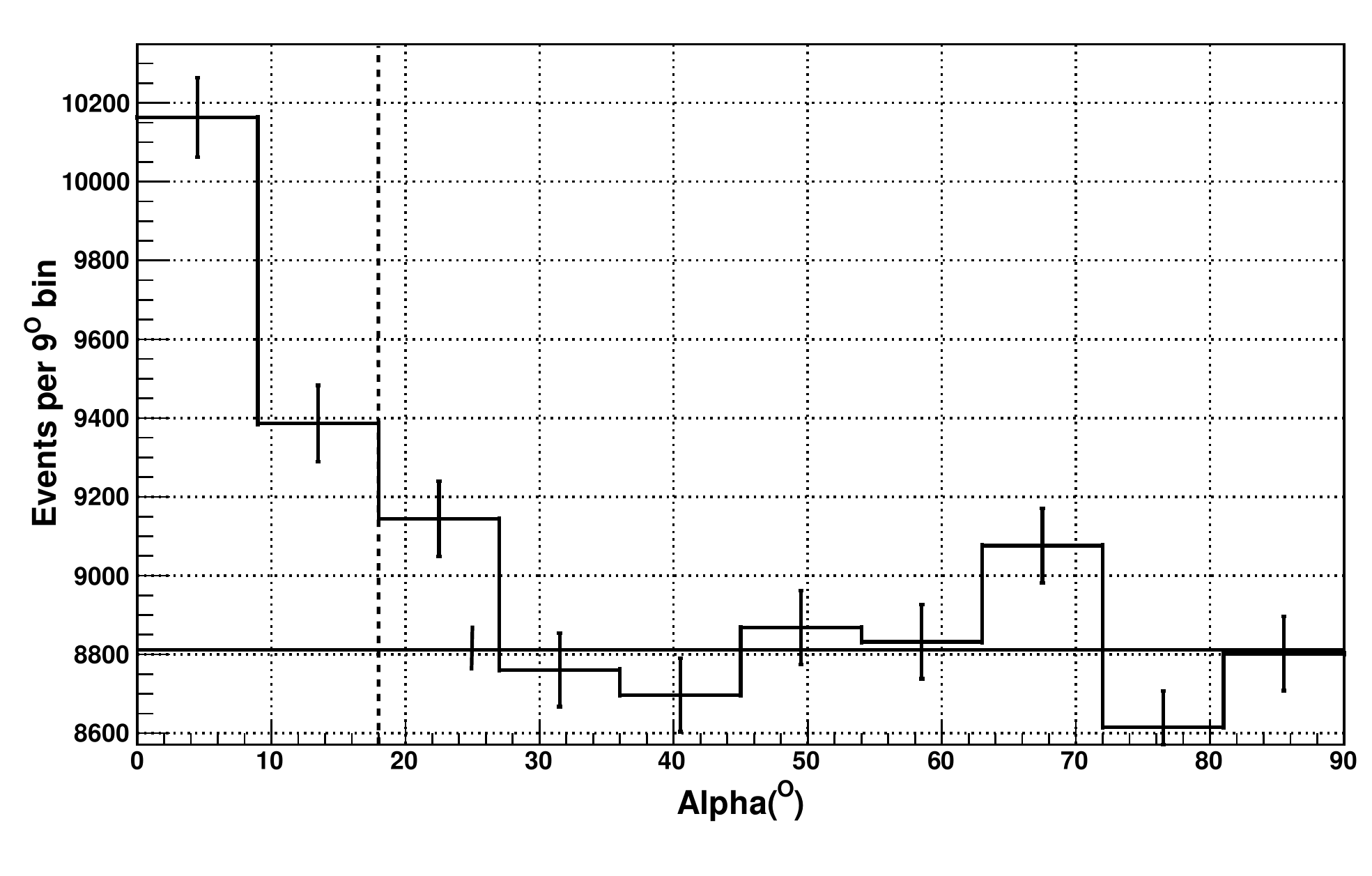}}
\caption{Distribution of image parameter $\alpha$ for  the  TACTIC observations  of Mrk 421  during  2009-10. Vertical dashed line represents the   $\alpha$ parameter cut used for  $\gamma$-ray like events.   Horizontal line represents the mean background level per 9$^{\circ}$ bin derived by using the  reasonably flat $\alpha$ region  of  27$^{\circ}$ $\leq$ $\alpha$ $\leq$ 81$^{\circ}$. Error bars shown  are for statistical errors only.}
\label{fig:alpha}
\end{center}
\end{figure}

Next, we present data analysis results of  Mrk 421 observations made  using the TACTIC telescope  during the year  2009-10. 
When all the  data   are analyzed together, the corresponding results obtained are shown in Figure \ref{fig:alpha}.  In this figure  the histogram  of the $\alpha$ parameter has been shown after having applied shape and orientation related imaging cuts given in Table \ref{tab:cuts}.  It  shows the presence of the VHE   $\gamma$-ray signal with a total number of $\gamma$-ray  like events  of 1932.7 $\pm$ 159.4.  The statistical significance of the detected signal  stands at 12.12$\sigma$ level, thereby indicating that  the source was   in a relatively high TeV emission state  during the  period of observations. 
When these data  were divided into seven monthly spells as depicted in Table \ref{tab:obs10} and  analyzed by using the same  data analysis procedure, the number of $\gamma$-ray like events  obtained   
%37.0 $\pm$ 20.2, 105.7 $\pm$ 38.4, 285.0 $\pm$ 62.7, 711.7 $\pm$ 88.6, 432.7 $\pm$ 88.6, 264.7 $\pm$ 52.9 and 96.0 $\pm$ 33.1 for  the respective spell, indicating that the source TeV $\gamma$-ray signal level 
was highest during February 2010 observations. The details of these results have been given in  Table \ref{tab:obs10}.
\\
%\vspace{0.5cm}

%\vspace{0.3cm}
\begin{table}[h]
\caption{ Monthly spell wise analysis of Mrk 421   data recorded during 2009-10 with only statistical errors.}
\begin{center}
% use packages: array
\begin{tabular}{|c|c|c|c|c|}
\hline
Spell  &  Observation Period & Time (hrs.) &  $\gamma$-rays & Significance\\ \hline
1 &  22-27 Nov. 2009  &  5.6  &     37.0 $\pm$ 20.2  &  1.83 \\ \hline
2 &  13-27 Dec. 2009  &  16.9 &     105.7 $\pm$ 38.4  & 2.75 \\ \hline
3 &  10-24 Jan. 2010  & 41.0  &     285.0 $\pm$ 62.7 &  4.54\\ \hline
4 &  06-23 Feb. 2010  &  50.2 &     711.7 $\pm$ 88.6  & 8.03 \\ \hline
5 &  07-22 Mar. 2010 & 59.4  &     432.7 $\pm$ 88.6 &  4.88 \\ \hline
6 &  02-18 Apr. 2010 & 38.5 &     264.7 $\pm$ 52.9 &  5.00 \\ \hline
7 &  02-16 May  2010  &  18.3 &     96.0 $\pm$ 33.1   &  2.90 \\ \hline
Total&  22 Nov. 2009-16 May 2010 &  229.9 &  1932.7 $\pm$ 159.4 &  12.12 \\ 
\hline 
\end{tabular}
\label{tab:obs10}
\end{center}
\end{table} 
\vspace{0.3cm}
\begin{figure}
\begin{center}
\mbox{\hspace{0cm}\includegraphics[scale=0.7, angle=90]{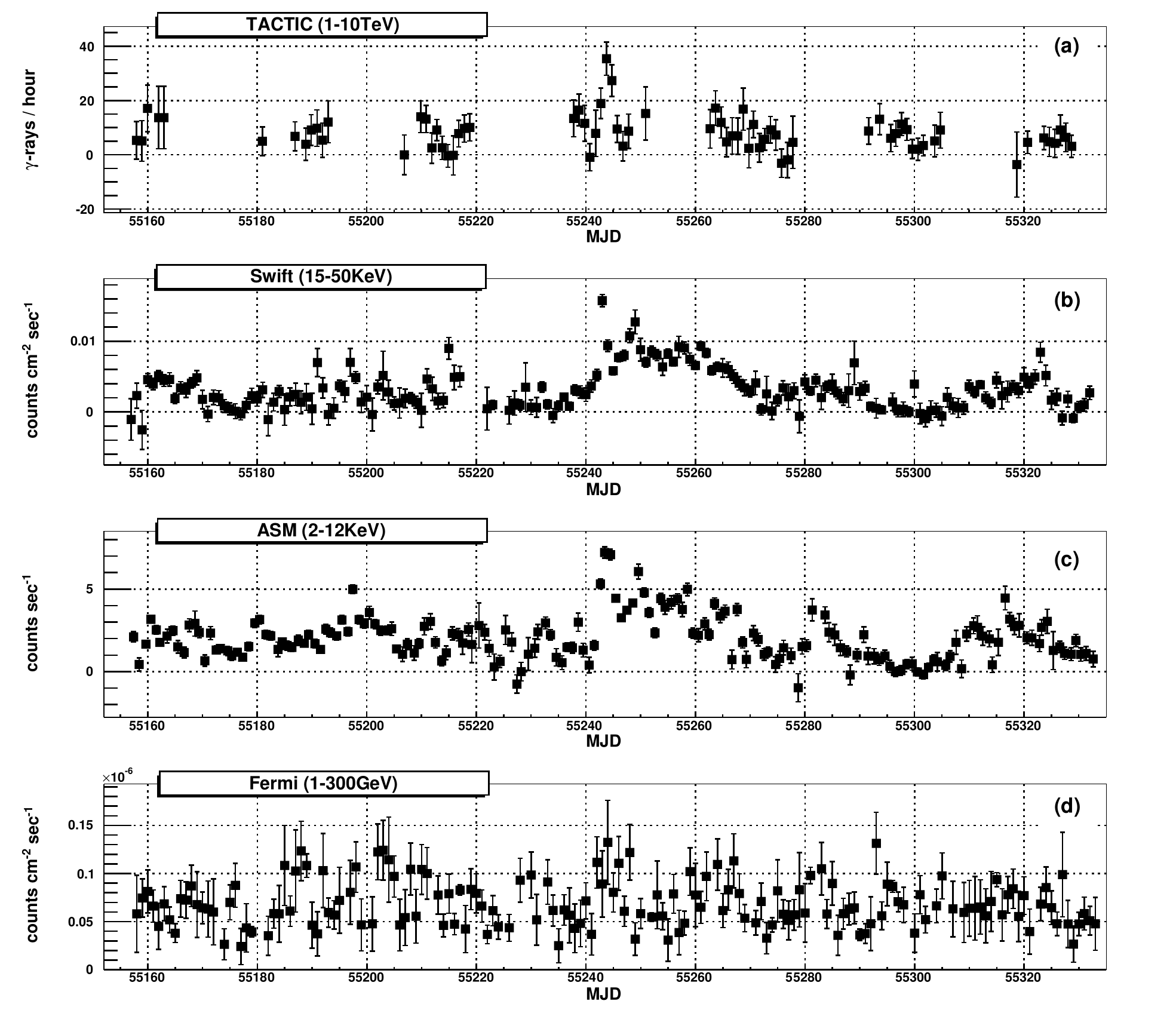}}
\caption{Mrk 421  light curves for 2009-10 TACTIC observations. }
\label{fig:lc10}
\end{center}
\end{figure}
%\vspace{0.5cm}
These data  have also been analyzed  on nightly basis in order to look out for  the possibility of  a strong episodic TeV emission. The  results so obtained are depicted in Figure  \ref{fig:lc10}a,  which shows a day-to-day variations of the $\gamma$-ray rate ($\gamma$-rays/hour) for Mrk 421 during 2009-10  observations. 
This light curve is characterised with  a reduced $\chi^2$ value of 83.26/69 with respect to the mean level of 7.4$\pm$0.65 photons per hour, with corresponding probability of 0.12 which is consistent with the no variability hypothesis.

Next we compare the TeV light curve obtained which is  shown in  Figure  \ref{fig:lc10}a  with those of 
the source with the  \textit{Swift}/Burst Alert Telescope (BAT) \cite{swift}( 15-50 KeV),  Rossi X-ray Timing Explorer (RXTE)/All-Sky Monitor (ASM)\cite{ASM}( 2-10 KeV) and \textit{Fermi}/Large Area Telescope (LAT)  \cite{fermi}(1-300 GeV) detectors. The \textit{Swift}/BAT contemporary light curve of the  source obtained from its archived data \cite{swift}  is  shown in Figure \ref{fig:lc10}b. It  is characterised with  a reduced $\chi^2$ value of 1594/168
 (probability is close to zero   which is consistent with the  variability hypothesis)
 with respect to the mean level of 0.00349 $\pm$ 7.4e-05 counts $cm^{-2}s^{-1}$.
 Next the RXTE/ASM
contemporary light curve shown in Figure \ref{fig:lc10}c which  has been plotted by using the daily average count rates of the ASM   from its archived data \cite{ASM}. The ASM light curve is characterised with  a reduced $\chi^2$ value of 2825/174 ( probability obtained is very low which  is consistent with the  variability hypothesis) with respect to the mean level of 2.15 $\pm$ 0.025 counts/sec.

The \textit{Fermi}/LAT
contemporary light curve shown in Figure \ref{fig:lc10}d   has been plotted by using the daily  counts $cm^{-2}s^{-1}$ of the LAT  from its archived data \cite{fermi}. The LAT light curve is characterised with  a reduced $\chi^2$ value of 291/156 ( probability obtained is very low which  is consistent with the  variability hypothesis) with respect to the mean level of 6.17e-08 $\pm$ 1.25e-9 counts $cm^{-2} sec^{-1}$.

\begin{figure}
\begin{center}
\mbox{\hspace{0cm}\includegraphics[scale=0.62,angle=90]{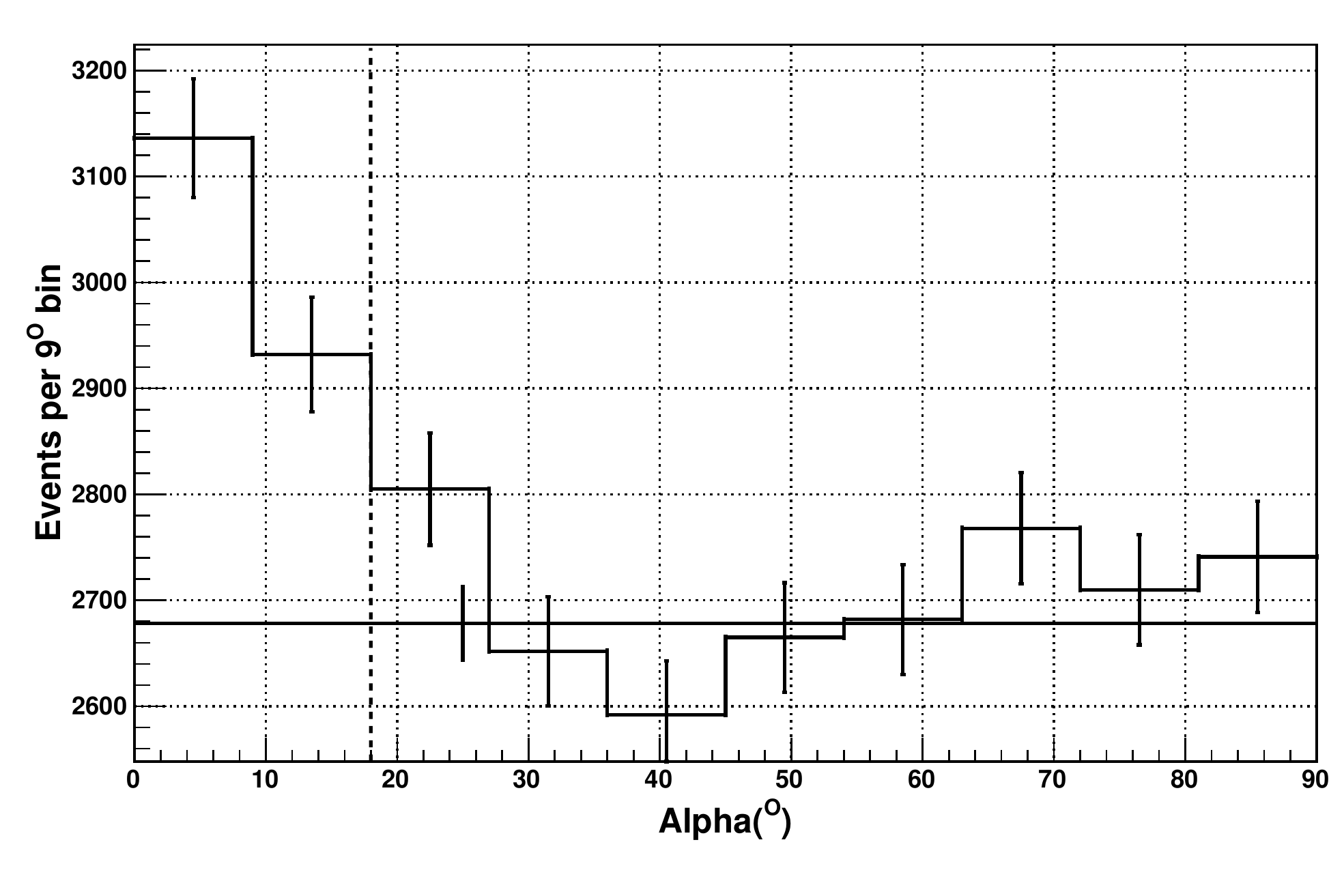}}
\caption{Distribution of image parameter $\alpha$ for  the  TACTIC observations  of Mrk 421  during  February 2010.Vertical dashed line represents the   $\alpha$ parameter cut used for  $\gamma$-ray like events.  Horizontal line represents the mean background level per 9$^{\circ}$ bin derived by using the  reasonably flat $\alpha$ region  of  27$^{\circ}$ $\leq$ $\alpha$ $\leq$ 81$^{\circ}$. Error bars shown  are for statistical errors only.}
\label{fig:alpha_feb}
\end{center}
\end{figure}
%\subsection{Analysis  and results of 2008 data}
\begin{figure}
\begin{center}
\mbox{\hspace{0cm}\includegraphics[scale=0.6]{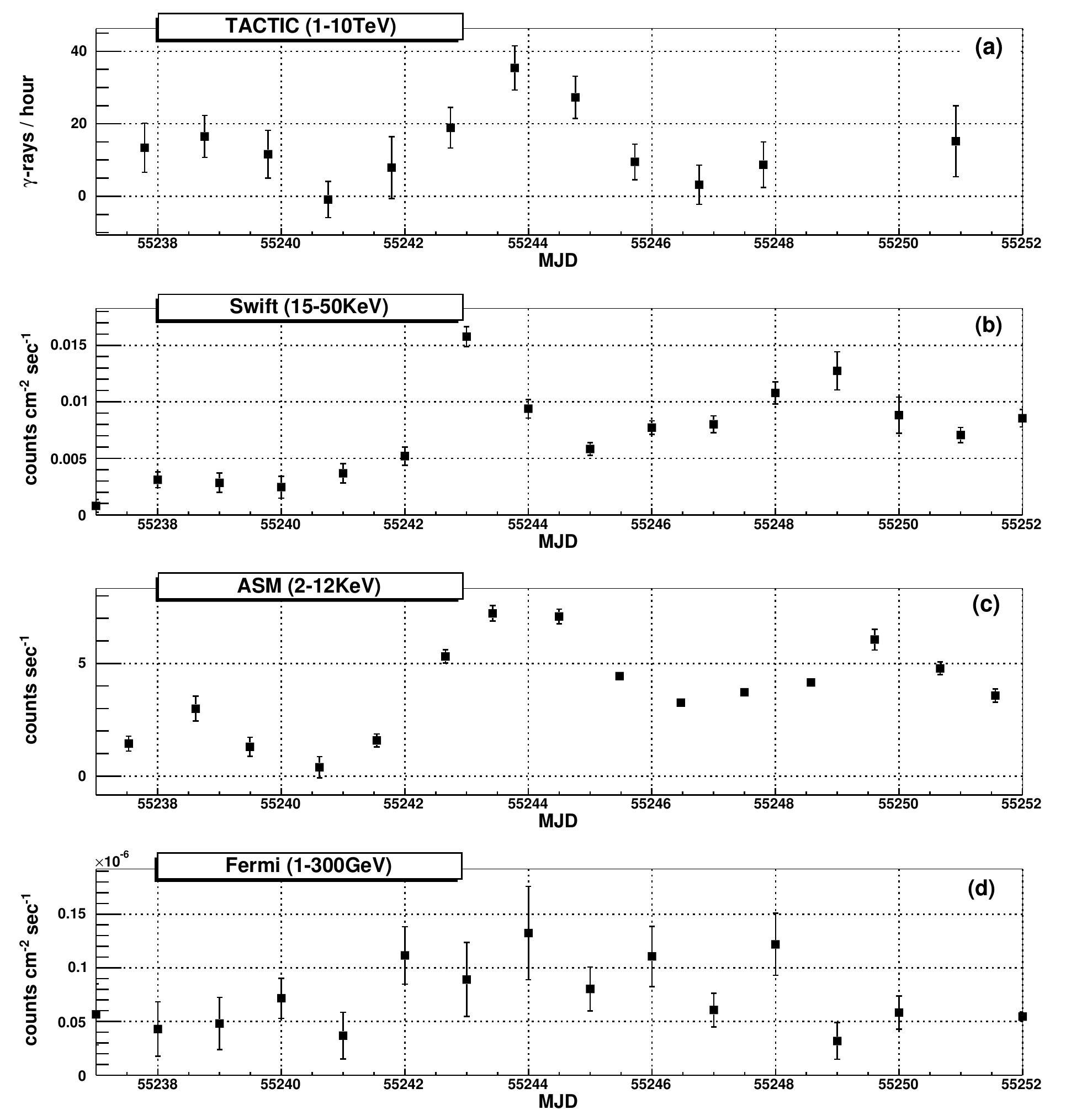}}
%\mbox{\hspace{0cm}\includegraphics[scale=0.7]{figs/fig3.eps}}
\caption{Mrk 421 light curves for February 2010  observations.}
\label{fig:lc_feb}
\end{center}
\end{figure}
As  already mentioned above and is also clear from  Table \ref{tab:obs10}  that  we have  detected a  statistically significant TeV $\gamma$-ray signal at 8.03$\sigma$ level  during  February TACTIC observations of the source. 
It was observed for 12 nights  and a total of  50.2 hours of on-source observations were possible during this month. Total number of TeV $\gamma$-rays detected during this month are 711.7$\pm$ 88.6 which is clear from the corresponding alpha plot shown  in  Figure \ref{fig:alpha_feb}. Further, we find that the source was at higher  TeV emission state during    three nights  of 15- 17 February, 2010(MJD 55242.73913-55244.98546). Details of these  results are given in Table \ref{tab:15_17feb}. On   16th February, 2010 we  detected  177.7 $\pm$ 30.8  $\gamma$-ray photons of TeV energies at more than 5$\sigma$ statistical significance level. During other two nights of 15th and 17th February, 2010 again we detected statistically significant number of TeV $\gamma$-rays  with greater than  4$\sigma$ level. Using these results we believe that the source  had possibly switched  to an intense flaring state during   15- 17 February, 2010  observations. 
\begin{table}[h]
\caption{Details of Mrk 421  data analysis results of TACTIC observations made during 15- 17 February 2010.}
\begin{center}
% use packages: array
\begin{tabular}{|c|c|c|c|c|c|}
\hline
Obs. Date &  Start MJD & End MJD &$\gamma$-rays & Sig.($\sigma)$ & Obs. Time(hrs)\\ 
(Feb. 2010)&&&&& \\ \hline
15 & 55242.73913      & 55242.99233      &     93.3 $\pm$ 27.5 & 3.39  & 4.9 \\ \hline
16 & 55243.77758      & 55243.98686     &     177.7 $\pm$ 30.8 & 5.76 & 5.0\\ \hline
17 & 55244.75948      & 55244.98546      &     143.0 $\pm$ 30.2 & 4.73 & 5.2\\ \hline
15, 16, 17 &  55242.73913     &  55244.98546 &    414.0 $\pm$ 51.2 & 8.09 & 15.1\\ 
\hline
\end{tabular}
\label{tab:15_17feb}
\end{center}
\end{table}
Further, in order to compare the February 2010 TeV results with those at lower energies, we have  depicted  the February source    light curves of TACTIC, \textit{Swift}/BAT, RXTE/ASM and \textit{Fermi}/LAT   in  Figures \ref{fig:lc_feb}a,b,c and d respectively. 
%along with those  at lower energies obtained by satellite based experiments Swift(BAT), RXTE(ASM) and Fermi(LAT)   in  Figures \ref{fig:lc_feb}b, c and d respectively.
As is clear from these figures  the daily  variations in the number of counts detected by these four experiments  are such that when we fit zero degree polynomials  to these light curves the reduced $\chi^{2}$ values obtained are 33.46/11, 353.8/15, 484.1/14  and  22.38/14 ($\chi^{2}$/dof )  for TACTIC, \textit{Swift}/BAT, RXTE/ASM and \textit{Fermi}/LAT respectively.
\textit{Fermi}/LAT's light curve reduced $\chi^{2}$ value of 22.38/14 yields a corresponding probability of 0.07 thereby  indicating that the light curve is consistent with  a flat distribution. However, the first three light curves  yield corresponding probabilities much less than 0.05   thereby indicating in favour of  variable signal detection in the   TACTIC, \textit{Swift}/BAT and RXTE/ASM energy ranges.  These light curves  also  indicate a possible  correlation between TeV and X-ray emissions from this source direction particularly during 15- 17 February, 2010  observations. 
This
lends   possible support to the SSC model in which a unique electron population produces the X-rays by
synchrotron radiation and the 
$\gamma$-ray component by inverse Compton scattering.
 %As mentioned earlier,  in   the  SSC model VHE $\gamma$-rays are produced by the inverse Compton scattering of low energy synchrotron photons by the same population of relativistic electrons in  jets   which produces  the first hump in the broadband SED of blazars, therefore a correlation in the X-ray and TeV $\gamma$-ray energies is expected. Hence   our these February, 2010  observations  of the source  possibly  support  the  model.  However, we do note that the above mentioned  correlation feature has already been observed earlier by  the  advanced ground based  $\gamma$-ray telescopes barring few orphan TeV flares. 

\begin{figure}[h]
\begin{center}
\mbox{\hspace{0cm}\includegraphics[scale=0.6]{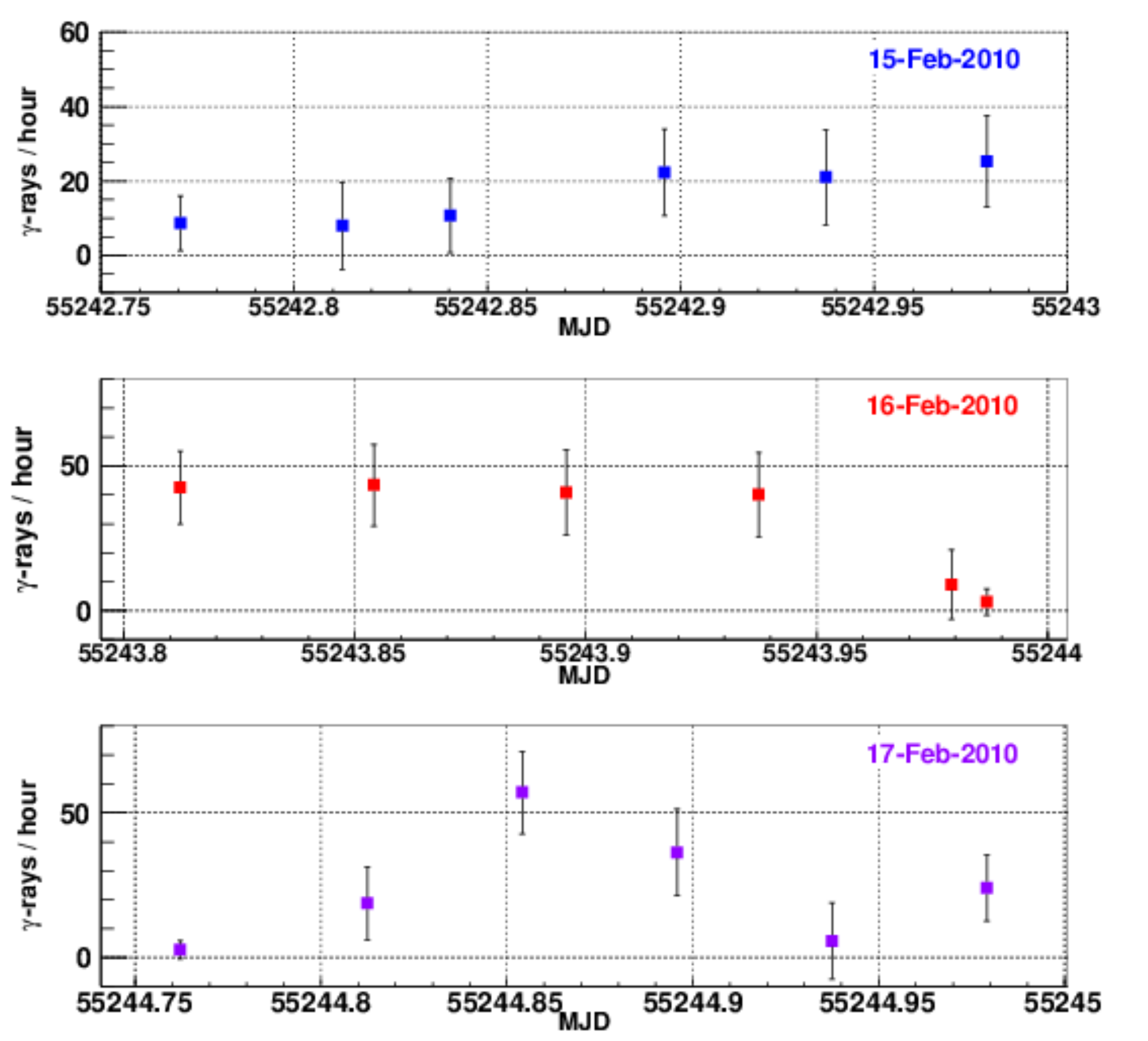}}
\caption{TACTIC observed hourly light curves of Mrk 421 on three nights of 15,16 and 17 February 2010 (MJD 55242.73913 to  55244.98546). %As is clear from the light curve of 17 February 2010, during the hourly run from MJD 55244.812502 to 55244.854161 we have  detected 57 $\pm$ 14.2  $\gamma$-ray photons of TeV energies  at 4$\sigma$ statistical significance.
}
\label{fig:lc_16}
\end{center}
\end{figure}
Further when hourly data runs  of these three days are investigated in terms of the detected number of $\gamma$-ray like events, we find that on the night of 17th February, 2010,  during the hourly run from MJD 55244.812502 to 55244.854161,  we have  detected 57 $\pm$ 14.2  $\gamma$-ray photons of TeV energies at 4$\sigma$ statistical significance as is clear from Figure \ref{fig:lc_16}. %  during an hour of observations. 
It may be noted here that TACTIC detects on an average 9 $\gamma$-ray like events in one hour of observations from the Crab nebula direction.  Therefore, we conclude that the source has possibly exhibited  yet another  short duration   high TeV $\gamma$-ray emission state during this period.
\vspace{0.3cm}
\begin{table}[h]
\caption{Differential energy spectrum data for Mrk 421 in 2009-10 with the TACTIC telescope. Only statistical errors are  given below.}\label{tab:421sp_10}
\begin{center}
\begin{tabular}{|c|c|c|}
\hline
\textbf{Energy}      &\textbf{Diff. flux  }    &\textbf{Statistical error in flux }\\
(TeV) &   photons cm$^{-2}$ s$^{-1}$TeV$^{-1}$ &  photons cm$^{-2}$ s$^{-1}$ TeV$^{-1}$ \\
\hline 
1.00 &  1.75 $\times$10$^{-11}$     &  6.78 $\times$10$^{-12}$           \\ \hline
1.50 &  5.45 $\times$10$^{-12}$      &   1.11 $\times$10$^{-12}$        \\ \hline
2.22 &  2.15 $\times$10$^{-12}$     &   3.15 $\times$10$^{-13}$        \\ \hline
3.32 &   7.46 $\times$10$^{-13}$    &    1.54 $\times$10$^{-13}$       \\ \hline
4.95 &   3.89 $\times$10$^{-13}$    &   7.98 $\times$10$^{-14}$        \\ \hline
7.38 &   1.85 $\times$10$^{-13}$    &   4.31 $\times$10$^{-14}$        \\ \hline
11.00 &  5.11 $\times$10$^{-14}$    &   2.59 $\times$10$^{-14}$       \\ \hline
16.44 &   1.01 $\times$10$^{-14}$     &  1.24 $\times$10$^{-14}$       \\ \hline
\end{tabular} 
\end{center}
\end{table}
\begin{figure}
\begin{center}
\mbox{\hspace{0cm}\includegraphics[scale=0.75,angle=90]{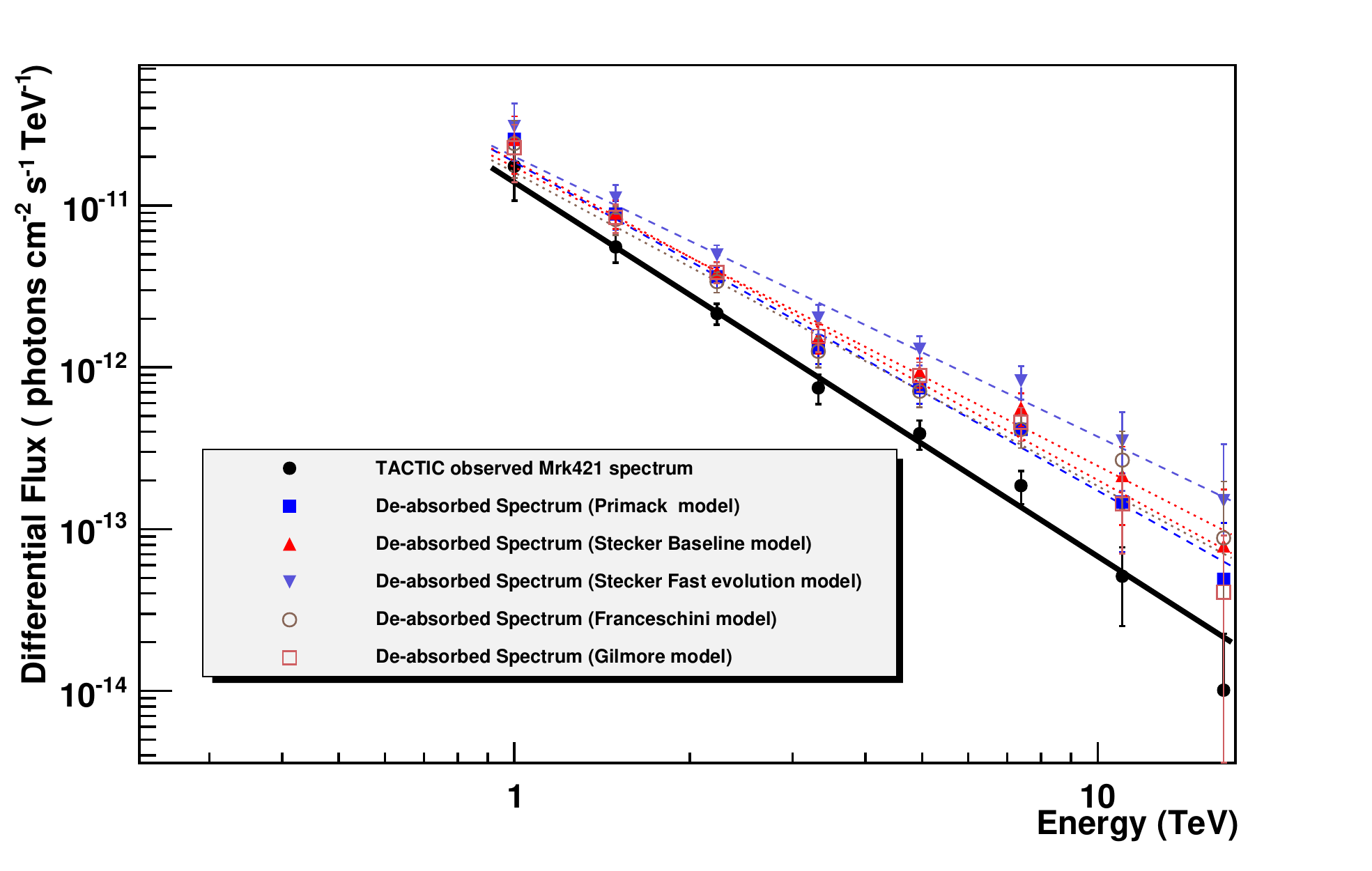}}
\caption{TACTIC observed differential energy spectrum of Mrk 421 derived using the 2009-10 observations( filled circle). De-absorbed source spectra estimated by using various EBL  models are also shown for comparison.  %(i)Primack given in \cite{pri05}(filled rectangle), (ii) baseline model given in\cite{ste07}(filled bottom base triangle ) and (iii) fast evolution model given in \cite{ste07}(filled top base triangle) (iv)Franceschini \cite{Franceschini08}( hollow circle) and (v)Gilmore  \cite{Gilmore09} (hollow rectangle).
} 
\label{fig:es}
\end{center}
\end{figure}
%\vspace{0.5cm}

\section{ Differential energy spectrum}

In this section we present the differential time-average energy spectrum of the source estimated  using its TACTIC observations made  during the  period   22 Nov. 2009- 16 May 2010.
The energy reconstruction procedure employed to unfold it  uses  
an Artificial Neural Network (ANN) with  3:30:1  configuration
and resilient backpropagation algorithm.  The energy of each  
 $\gamma$-ray like event is estimated  on the
basis of its image SIZE, DISTANCE and zenith angle. This  
method yields an energy resolution of $\sim 26\%$,   the details of which
 have been given %used to derive the differential time-average energy spectrum of  Mrk 421 are given
  in \cite{dhar09,ya07}.

% We have used 2009-10 data of $\gamma$-ray like events  to determine 
 The  observed Mrk 421 differential energy spectrum so obtained  has been tabulated in Table \ref{tab:421sp_10} and also shown in the  Figure \ref{fig:es} which  extends up to  16.44 TeV.  % with a power-law index of 2.31 $\pm$ 0.14.
%The  observed differential energy  spectrum of the source obtained  by  using the combined 2009-10  data sets is shown in the  Figure \ref{fig:es}. 
 The errors in the flux constant and   spectral index  are   standard errors. 
A power law  function of the form  $dF/dE=f_0 E^{-\Gamma}$ when  fitted to the data in the energy range 1- 16.44 TeV yields 
\begin{equation}
\frac{dF}{dE} =  (1.39 \pm 0.239)\times 10^{-11}cm^{-2}s^{-1}TeV^{-1}(\frac{E}{1.0TeV})^{-2.31\pm 0.14}
\end{equation}
for the observed differential energy spectrum of the source with
%$f_0=(6.8\pm1.4)\times 10^{-11}cm^{-2}s^{-1}TeV^{-1}$ and  $\Gamma=3.32\pm0.22$ with a
 $\chi^2/dof= 3.37/6$ (probability 0.76).
\vspace{0.3cm}

\begin{table}[h]
\caption{Estimated intrinsic differential energy spectrum fitted parameters for three models. }\label{tab:models}
\begin{center}
\begin{tabular}{|c|c|c|c|c|}
\hline
$\Gamma$ &  $\Delta\Gamma$   &    f$_{0}$   &       $\Delta$f$_{0}$   &         Model\\ \hline
 2.31	&  0.14&	 1.39 $\times$ 10$^{-11}$&	 2.39 $\times$ 10$^{-12}$&		Presented observations       \\  \hline
 2.03  &0.16 &	 1.86  $\times$ 10$^{-11}$	 & 3.60 $\times$ 10$^{-12}$ 	&	Primack \cite{pri05}     \\  \hline
 1.85 & 0.16	& 1.73 $\times$10$^{-11}$&	 3.16 $\times$ 10$^{-12}$ 		&	Stecker Baseline   \cite{ste07}   \\  \hline
 1.73&	  0.17	& 2.00 $\times$10$^{-11}$&	 4.11 $\times$ 10$^{-12}$ 		&	Stecker Fast evolution   \cite{ste07}   \\  \hline
 1.94&	  0.18	& 1.60 $\times$10$^{-11}$&	 3.43 $\times$ 10$^{-12}$ 		&	Franceschini  \cite{Franceschini08}    \\  \hline
 1.97&	  0.14	& 1.88 $\times$ 10$^{-11}$&	 3.20 $\times$ 10$^{-12}$ 		&	Gilmore   \cite{Gilmore09}   \\  \hline
\end{tabular}
\newline\newline
\end{center}
\end{table}

%  TACTIC detected February flare

%==================================
We have also  estimated the source intrinsic $\gamma$-ray spectra by using the SED of the EBL provided by   five models given in Table    \ref{tab:models}.
% \cite{pri05}, \cite{ste07}
%(Primack, baseline and fast evolution models),
%,\cite{Franceschini08} and  \cite{Gilmore09}. 
The resulting intrinsic spectra for  five models are shown in Figure \ref{fig:es}.  A power law fit to the estimated intrinsic data of the type $d\Phi/dE=f_0 E^{-\Gamma}$ in the energy range 1- 16.44 TeV yields values of $\Gamma$ and $f_{0}$ listed in  Table \ref{tab:models}. As is clear from  Figure \ref{fig:es}, the Primack model yields minimum attenuation of TeV photons, in contrast,   Stecker's fast evolution model yields maximum attenuation  as compared to other models used in estimating the intrinsic source spectra. 

As  mentioned earlier, Mrk 421 is a nearby source with z=0.031, so we do not expect  significant absorption of intrinsic source spectrum due to EBL in the TACTIC energy range. Another important characteristic feature associated with this source is its relatively higher  flaring frequency  when compared to other objects of the same class, thereby making it an excellent cosmic laboratory to study VHE $\gamma$-ray emission mechanisms and physics of relativistic jets.

\begin{comment}
\begin{figure}
\begin{center}
\mbox{\hspace{0cm}\includegraphics[height=10.0cm]{pic/figure7.pdf}}
\caption{Correlation between (a) TeV and RXTE(ASM) (b)TeV and Swift(BAT)  (c) TeV and Fermi(LAT) 
observations. The maximum allowed time difference between the data points $\Delta$t $<1$  days. }
\label{fig:cor}
\end{center}
\end{figure}
\end{comment}
\section{Discussion and conclusions}

\begin{figure}
\begin{center}
\mbox{\hspace{0cm}\includegraphics[height=14.0cm,angle=90]{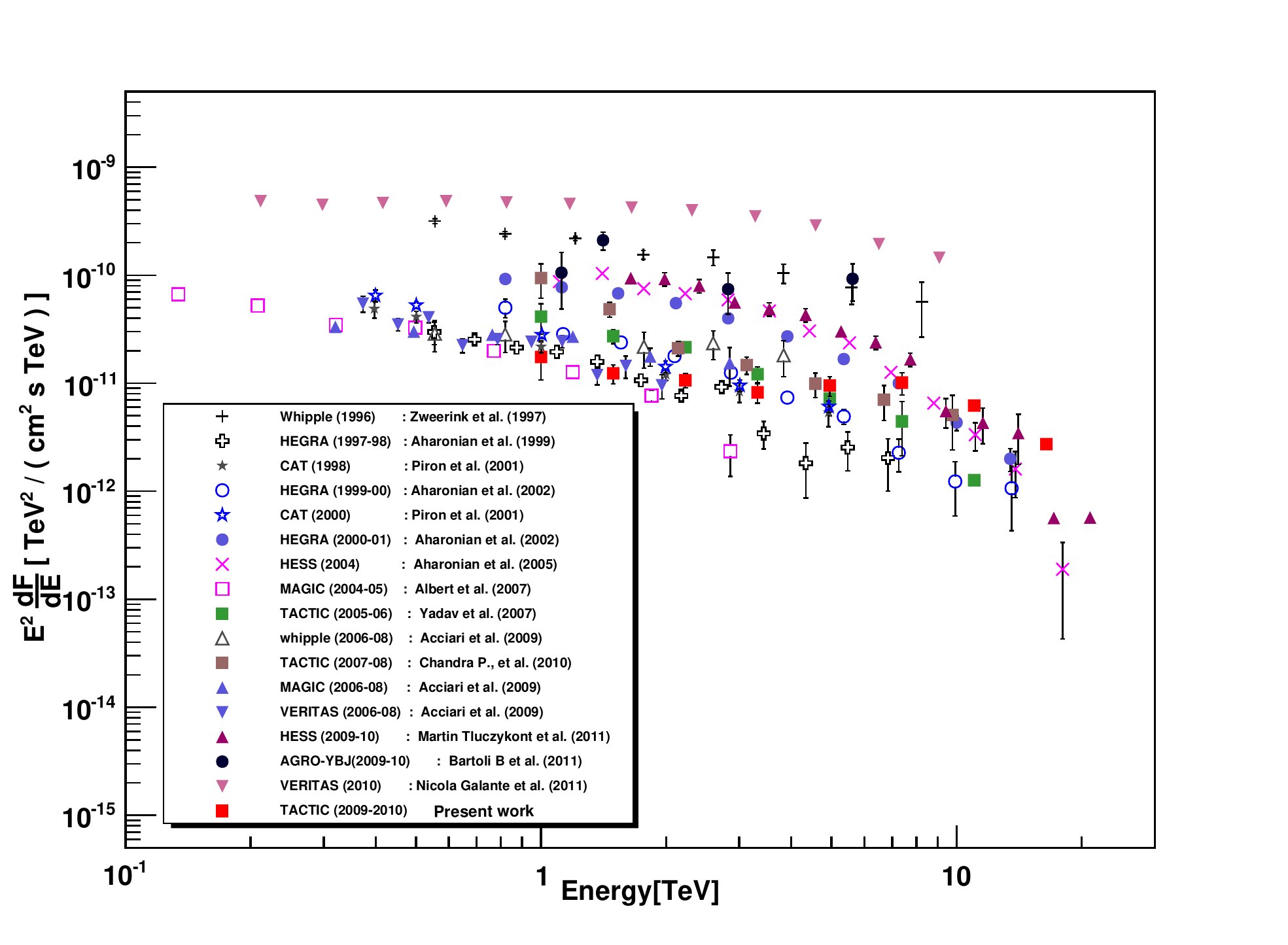}}\\

\caption{Comparison of the  observed  spectral energy   distribution  of Mrk 421 in the VHE range obtained by
various groups.
% the HESS, MAGIC, VERITAS, Whipple, HEGRA, CAT, ARGO-YBJ and TACTIC  \cite{ah05,al06,zw97,acc09,ah99} \cite{ah02,pir01,ya07,he11,Mart11}. 
TACTIC observed differential energy spectrum during 2009-10 observations is in the energy range 1.0 - 16.44 TeV. 
}
\label{fig:sed}
\end{center}
\end{figure}
\vspace{0.5cm}
For the reasons mentioned above, we are monitoring Mrk 421 on the long term basis with TACTIC $\gamma$-ray telescope.
In this paper, we have presented the  results obtained using TACTIC   observations  made during 2009-10.  Detailed analysis of this data shows  evidence for the  presence of a statistically significant VHE $\gamma$-ray signal at  12.12$\sigma$ level. The total number of $\gamma$-ray like events detected are 1932.7 $\pm$ 159.4.  
These  observations  suggest that the source had  gone into a   variable active  state in the VHE region during 
the period of TACTIC 2009-10 observations. 
During the month of February 2010, TeV signal strength in terms of the number of TeV $\gamma$-ray like events which is   711.7$\pm$ 88.6 highest as compared to results obtained for other  months as depicted in Table \ref{tab:obs10}. It is interesting here to mention that the HESS \cite{Mart11}, VERITAS \cite{nico11} and ARGO-YBJ \cite{Bart11,he11} have also detected the source in a high state 
particularly during February 2010.

 We have  estimated an observed   differential energy  spectrum of the detected $\gamma$-ray  like events which  is shown in Figure \ref{fig:es}. 
The spectrum  fits well with the power law function of the form ($dF/dE=f_0 E^{-\Gamma}$) with $f_0=(1.39 \pm 0.239)\times 10^{-11}cm^{-2}s^{-1}TeV^{-1}$ and  $\Gamma=2.31\pm0.14$. 
In comparison with our earlier TACTIC detected high states of this source reported in  \cite{ya07,chandra10}, present observed spectrum is harder   and extends up to 16 TeV primary $\gamma$-ray photon energy. 
We have also used five  templates of EBL 
 %given in \cite{pri05},  \cite{ste07}, \cite{Franceschini08} and  \cite{Gilmore09}
to estimate $\tau$'s for z = 0.031 which have been used to estimate corresponding  intrinsic source spectra.  The results of this study have been tabulated in Table \ref{tab:models}.

 A comparison of the  observed  spectral energy   distribution  of this source  in the VHE range obtained by the HESS, MAGIC, VERITAS, Whipple, HEGRA, CAT, ARGO-YBJ and TACTIC  \cite{ah05,al06,zw97,acc09,ah99} \cite{ah02,pir01,ya07,he11,Mart11} is shown in Figure \ref{fig:sed}. As is clear from this figure  our 2009-10 results are in close agreement with various reported results on this source.
 
TACTIC, \textit{Swift}(BAT) and RXTE(ASM) light curves of 15- 17 February, 2010  observations (55242.73913 - 55244.98546 )  do indicate  a possible correlation of TeV and X-ray emissions of the source as is shown in Figures  \ref{fig:lc_feb}a,b and c. This indicates that the mentioned short TeV $\gamma$-ray flare was not an orphan type as it is also seen at lower energies  and  possibly supports  the SSC model in which a relativistic  electrons population produces the X-rays by synchrotron radiation and the  $\gamma$-ray component by inverse Compton scattering.
%We plan to monitor this source on long term basis so as to study its energy spectra corresponding to various high TeV emission states with the TACTIC $\gamma$-ray telescope.  

\section{Acknowledgements} 
Authors would like to acknowledge the excellent team work of the technical staff at the  Mt. Abu observatory. We also thank the anonymous referees for their useful comments.

\end{document}